\documentstyle[12pt]{article}
\renewenvironment{thebibliography}[1]
 {
   \begin{list}{\arabic{enumi}.}
    {\usecounter{enumi} \setlength{\parsep}{0pt}
     \setlength{\itemsep}{3pt} \settowidth{\labelwidth}{#1.}
     \sloppy
    }}{\end{list}}
\def \MSbar {\vbox{\hrule\kern 1pt\hbox{\rm MS}}}
\def \GeV { {\ \rm GeV} }

\font\bigrm=cmbx12 scaled 1200
\font\titlefont=cmbx12 scaled 1200
\def \DESepsf(#1 width #2){\bf #1  here: just uncomment the macro.}
\begin{document}
%

\setcounter{page}{1}
\rightline{OITS 528}
\vskip -3 pt
\rightline{29 November 1993}
\vskip -3 pt
\vglue 0.5 in
\centerline{\titlefont MULTIPARTICLE DYNAMICS FROM 1983 TO 1993}
\vskip 0.2 in
\centerline{\bf Davison E. Soper}
\centerline{\it Institute of Theoretical Science}
\centerline{\it University of Oregon, Eugene, OR  97403, USA}
\vskip 0.3 in
\centerline{Talk at the XXIV International Symposium
on Multiparticle Dynamics}
\centerline{Aspen, Colorado, September 1993}
\vskip 0.3 in
\centerline{ ABSTRACT }
\vskip 0.2in
\baselineskip=14truept plus 0.2truept minus 0.2truept

\parbox{6in}{
I compare our understanding selected topics in Multiparticle Dynamics
at this meeting to what we knew at the 1983 Multiparticle Dynamics
Symposium.  I also discuss rapidity gap physics, a subject that has
developed in the years since 1983.}

\vglue 1.0cm
{\bigrm\noindent 1. Introduction}
\vglue 0.4cm
\baselineskip=16truept plus 0.2truept minus 0.2truept

Ten years ago the XIV International Symposium on Multiparticle
Dynamics took place at Lake Tahoe, California. In this year of 1993, we
are met again at a beautiful mountain setting: Aspen, Colorado. At the
1983 conference several questions concerning the dynamics of
elementary particles were of particular concern, and the participants
looked to the future for a better understanding. That future is now.
It thus seems a good opportunity to assess what answers we have found
and what developments have played a major role in whatever progress has
been achieved.  I am afraid that I cannot project ten years into the
future, particularly as the fate of the Superconducting Super Collider
hangs in the balance, but we may be able to at least discern some
directions along which the research frontier is yielding to determined
efforts.  I will concentrate on theoretical progress, but experimental
developments will be necessarily intertwined with the story. Toward the end of
the talk I will turn to the newly emerging
question of rapidity gap physics, a question that was not contemplated
in 1983.

I should point out that this talk will be a selection of just a few
topics that seemed to me most interesting to discuss today, rather
than a comprehensive review.  I would have liked, for instance, to
discuss the partonic structure of the photon, since this was an area
of considerable interest at the 1983 conference and will, I think, be
an area of substantial research progress in the next few years as a
result of HERA experiments. Similarly, I would have liked to discuss
quark-gluon plasma formation in heavy ion collisions and its
subsequent decay, possibly with temporary misalignments of the chiral
pion and sigma fields from their vacuum values \cite{alaska}. The
reader will undoubtedly find some other favorite topic missing, but I
hope that enough is left to capture his or her interest.

\vglue 0.6cm
{\bigrm\noindent 2. Beyond the Standard Model physics}
\vglue 0.4cm

The 1983 conference featured an excellent talk by J. Ellis
\cite{jellis} in which he advocated viewing multiparticle dynamics not
as ``an end in itself, but as a means for advancing to the next stage
in physics.''  He pointed out that ``the Standard Model with its
elementary Higgs fields may appear satisfactory at first sight, but it
has problems'' associated with its lack of explanation of why the mass
scale of the Higgs sector is less than a TeV rather than the Grand
Unified theory scale of some $10^{15} \GeV$ or the Plank scale of
$10^{19} \GeV$.  Ellis's favorite candidate for a theory beyond the
Standard Model that solves this hierarchy problem was supersymmetry
(SUSY).  He urged the importance of searching for the expected
supersymmetric partners of the quarks, leptons, gauge bosons, and
Higgs bosons of the Standard Model.

Where are we today? We haven't found SUSY. However, good chunks of the
mass ranges for the expected particles have been explored.  I would
say that, given the negative results of this exploration, the
prospects for supersymmetry being nature's solution of the hierarchy
problem are greatly diminished.  I would say that, except that in ten
further years of invention, we theorists have not been able to come
up with a credible solution for the problem other than
supersymmetry.  Furthermore, during the last ten years, an
exciting candidate for a theory of everything has emerged, the
superstring theory. One cannot put much faith in this theory until it
is well enough developed to produce quantitative predictions.  However
the case for supersymmetry is strengthened by the fact that
supersymmetry is a natural feature of superstring theory. Another
hint in favor of supersymmetry comes from grand unified gauge
theory.  In such a theory, the running values of the three couplings
$g_i(\mu)$ of the Standard Model should  reach the same value $g$ at
the same value of the scale parameter $\mu$.  With recent
improvements in our knowledge of the value of the strong coupling
$g_s(M_Z)$ from LEP, one can now see that this condition is {\em not}
met by the Standard Model, but is met if the Standard Model is
supplemented by supersymmetry with a symmetry breaking scale in a
reasonable range, about 1 TeV \cite{amaldi}. So where are we today? We
are still confused, but at a higher level.

\vglue 0.6cm
{\bigrm\noindent 3. Electroweak Standard Model physics}
\vglue 0.4cm

I now turn to Standard Model dynamics.  1983 was a watershed year for
the Standard Model.  On January 21 and 22, 1983, at seminars at the
main auditorium at CERN, the UA1 and UA2 groups reported the
experimental discovery of the $W$ boson \cite{nobel}.  The discovery
of the $Z$ was announced by UA1 at a CERN seminar on 27 May and in a
paper submitted to Physical Review Letters shortly thereafter.  The
accumulated evidence from UA1 was presented at the 1983 Multiparticle
meeting in June in an exciting talk by E.~Locci \cite {locci}. There
were 52 $p\bar p \to W + X \to e + \nu+ X $ events, 4  $p\bar p \to Z
+ X \to e^+ + e^- + X $ events and 2 $p\bar p \to Z + X \to \mu^+ +
\mu^- + X $ events. From these, UA1 obtained the values shown in
Table~1 for $M_Z$, $M_W$, and  $\sin^2(\Theta_W)$, in good agreement
with the expectations based on the Standard Model and low energy
electroweak experimental results.  They obtained $\rho \equiv
M_W^2/(M_Z^2 \cos^2(\Theta_W)) =0.925 \pm 0.05$, in agreement with
the Standard Model value $\rho = 1$.

\begin{table}[htb]
\begin{center}
\begin{tabular}{|l|c|c|}\hline
&1983&1993\\\hline
$M_W$ (GeV) \rule{0mm}{4mm}& 80.9$\pm$3.4 & 80.47$\pm$0.3\\\hline
$M_Z$ (GeV) \rule{0mm}{4mm}& 95.6$\pm$3.3 & 91.187$\pm$ 0.007\\\hline
$\sin^2(\Theta_W)$\rule{0mm}{4mm} & 0.226$\pm$0.016 & 0.2321$\pm$
0.0006\\\hline
\end{tabular}
\end{center}
\caption{Measured value of electroweak parameters in 1983 and 1993. }
\end{table}

Measurements at LEP over the past four years have enormously increased
the precision with which we have tested the electroweak sector of the
Standard Model.  This is illustrated by the 1993 values for $M_Z$ and
$\sin^2(\theta_W)$ shown in Table~1, taken from the review of LEP
results by  S.~L.~Wu at this conference \cite{wu}.  I also show the
1993 value for $M_W$, obtained from $\bar p p$ experiments at Fermilab
\cite{Wmass}.  You can see that the 1983 results were correct, within
their errors.  But what is really remarkable is how far the errors
have been reduced.  For $M_W$ the error has been reduced by a factor
10 (with further reductions in sight).  For $M_Z$ the error has been
reduced by a factor of 400 and for $\sin^2(\theta_W)$ by a factor of
30.

This improvement in precision is largely a story of accelerators and
careful experiments, but substantial theoretical work has been
required and has been carried out in order to match the experimental
precision.  We are now testing the theory at the one loop level. One
significant result is that we can now give a value, with
errors, for the top quark mass even though the top quark has not been
directly seen.  The value reported at this conference was $M_t =
166^{+17+19}_{-19-22}\GeV$ \cite{wu}.  Another significant result is
that one can constrain the masses and couplings of possible new
particles that might contribute to electroweak loop diagrams.

\vglue 0.6cm
{\bigrm\noindent 4. Focus on the strong interactions}
\vglue 0.4cm

In their Preface to the 1983 Proceedings \cite{gunion}, P.\ Yager and
J.\ F.\ Gunion wrote

\begin{quote}
The symposium focused on the implications on tests of of quantum
chromodynamics (QCD) over the full range of interactions: from low
$p_T$ to high $p_T$; from $e^+e^-$ collisions to $\bar p p $
collisions; and from simple hadron-hadron collisions to nucleus-nucleus
collisions.  A principle aim of the conference was to increase our
understanding of the extent to which non-perturbative effects such as
confinement and final state fragmentation can, firstly, be isolated
from hard processes so as to test perturbative QCD, and, secondly, be
understood on the basis of QCD.\dots

Indeed, if one overall conclusion can be drawn from the conference,
it is that multiparticle dynamics must be still better understood
both phenomenologically and theoretically before it will be possible
to test in detail the correctness of QCD, either perturbatively
or non-perturbatively.
\end{quote}

I shall devote the next sections to a discussion of how
well the aims articulated by Yager and Gunion have been realized
during the past ten years.  Briefly, it appears to me that we have
done rather well on testing the correctness of QCD.  We have done less
well in understanding non-perturbative multiparticle phenomena on the
basis of QCD, but I think there has been some progress.

\vglue 0.6cm
{\bigrm\noindent 5. QCD in $e^+e^- \to$ hadrons}
\vglue 0.4cm

At the 1983 Symposium, G.~Wolf reviewed studies of the strong
interactions in  $e^+e^-$ annihilation experiments at PETRA and PEP
\cite{wolf}. In particular, he focussed on the determination of
$\alpha_s$ from these experiments.  There is certainly more to
testing QCD than simply measuring $\alpha_s$ in various ways and
seeing if you get the same results within the errors.  However, a
comparison of such measurements in 1983 and 1993 can serve as an
indicator of the development of the state of the art between these
two times.

In Table~2, I show a selection of the results for $\alpha_s$ reported
in Wolf's talk. In each case I have used the renormalization group to
translate from $\alpha_s(34 \GeV)$ to $\alpha_s(M_Z)$. First is the
value obtained from the total hadronic cross section.  The theory for
this quantity is simple, but the experimental errors are large.  (This
is because $\sigma_T \propto 1 + \alpha_s / \pi + \cdots$, so that a
2\% measurement of $\sigma_T$ yields a 30\% measurement of
$\alpha_s$). Next I show the values obtained from various measures of
the distributions of hadrons in the final state, {\it i.e.} the shapes
of the events.  The results depend on the fragmentation model used,
either the independent fragmentation of each jet or the string model.
These results were from the TASSO group.  Finally, I show an analysis
from the CELLO group based on the asymmetry of the energy-energy
correlation function, which tells how the energy going in one
direction is correlated with the energy going in a different direction
as a function of the angle between the two directions.  Again the
results depend on which fragmentation model was used in the analysis.

\begin{table}[htb]
\begin{center}
\begin{tabular}{|l|c|}\hline
$\sigma_T(e^+e^-\to{\rm hadrons})$
&0.15 $\pm$ 0.05\\\hline
shapes, independent jets
&0.133 $\pm$ 0.01\\\hline
shapes, string fragmentation
&0.165 $\pm$ 0.01\\\hline
AEEC, independent jets
&0.105 $\pm$ 0.01\\\hline
AEEC, string fragmentation
&0.119 $\pm$ 0.01\\\hline
\end{tabular}
\end{center}
\caption{Measured values of $\alpha_s(M_Z)$ in 1983
\protect\cite{wolf}. }
\end{table}

In Table~3, I show values of $\alpha_s(M_Z)$ obtained by several
different methods as of 1993.  The values are taken from the review of
I.\ Hinchliffe at the 1993 Rencontre de Moriond \cite{hinchliffe},
except for the value for $\sigma_T(Z\to{\rm hadrons})$, which changed
significantly between spring and summer 1993 and which I took from the
talk of M.~Shapiro at the 1993 Lepton Photon Conference
\cite{shapiro}. Notice that there are a variety of  methods used and
that, in contrast to the situation in 1983, the values obtained using
these methods agree within their errors.

\begin{table}[htb]
\begin{center}
\begin{tabular}{|l|c|}\hline
$\sigma_T(Z\to{\rm hadrons})$
&0.122 $\pm$ 0.007\\\hline
$\sigma_T(\tau\to{\rm hadrons})$
&0.121 $\pm$ 0.011\\\hline
deeply inelastic scattering
&0.113 $\pm$ 0.006\\\hline
$Z\to{\rm hadrons}$\ event shapes
&0.121 $\pm$ 0.008\\\hline
$\Gamma(\Upsilon\to{\rm hadrons})$
&0.108 $\pm$ 0.010\\\hline\hline
Average
&0.119 $\pm$ 0.005\\\hline
\end{tabular}
\end{center}
\caption{Measured values of $\alpha_s(M_Z)$ in 1993. }
\end{table}

Considering just our example of $e^+e^-$ annihilation, why do we seem
to be doing better in 1993 than we were in 1983? One problem, which in
fact received a lot of discussion in 1983, was the effect of
non-perturbative, long-distance physics.  This problem is reflected in
the difference between the the independent jet model and the string
model results.  Partly we are better off now because of the higher
$\sqrt s$, 91 GeV versus 34 GeV on average.  This makes the
long-distance effects smaller (by a factor 3 or 9 depending on
whether the effects go like $m/\sqrt s $ or $m^2/ s $ ).  Partly we
are better off now because of better Monte Carlo event generators, of
which I will speak in a moment. Mostly, I think we are better off
because of a better theoretical framework and better calculations.

{\bf Calculations.} A serious confrontation between theory and
experiment requires next-to-leading-order calculations (otherwise the
theoretical errors are big and $\Lambda_{\rm QCD}$ isn't really
defined).  These were becoming available by 1983, but there were
disagreements among them.  By now, there is a unified
next-to-leading-order calculation by Z.~Kunszt and P.~Nason of all the
suitable distributions describing the final state in $e^+e^-$
annihilation \cite{kn}. This calculation, based on the 1980 matrix
elements of R.\ K.\ Ellis, Ross and Terrano \cite{ert}, agrees with
most of the special cases in the previous literature.

{\bf Theoretical framework.} In 1983 the difference between a measured
quantity that is infrared safe (like the thrust distribution) and a
quantity that is not (like the sphericity) was not understood. The
essential idea had been introduced in specific cases in 1977 and
1978 by Sterman and Weinberg, by Basham, Brown, S.\ D.\ Ellis, and
Love, and by Fox and Wolfram \cite{irsafe}.  But the
general principle had not really been absorbed. The quantities that
are not infrared safe are sensitive to long distance effects like the
splitting or joining of two collinear partons.  They should not be
used as a tool for examening short distance physics.  Indeed, if you
use  next-to-leading-order perturbation theory to calculate an
infrared unsafe quantity, you should get $\infty$.  To avoid the
infinities, you need to somehow insert a long distance model, and the
result will depend on the model. In 1983, this was not understood, and
infrared safe quantities were mixed with infrared unsafe quantities in
the analysis.

Looking back on the 1983 results, one may speculate that it was not
accidental that when an infrared safe quantity, the asymmetry of the
energy-energy correlation function, was  used in conjunction with the
better fragmentation model (the string model), the result obtained for
$\alpha_s$ was in good agreement with the present best value.

\vglue 0.6cm
{\bigrm\noindent 6. Improvements in Monte Carlo event generators}
\vglue 0.4cm

We have just seen an example in which Monte Carlo event generators
have made a real difference.  Such generator codes are now regarded as
an essential part of particle physics experimentation and analysis.
They were emerging as important tools around 1980 and advanced
tremendously in power and usefulness in the period 1983 to 1993. Among
the heros of this story are Paige and Protopropescu, Gottschalk,
Andersson, Gustafson, Sjostrand, Ingelman, and Webber and Marchesini.
Let me mention two of the important technical features that were
introduced or received widespread implementation during this period
\cite{montecarlo}.

{\bf Hadronization schemes}. Two hadronization schemes are commonly
used in the modern Monte Carlos. In the string picture, one imagines
that color flux tubes (the strings) join outgoing partons; these
strings then break into pieces with quarks and antiquarks at the newly
formed string ends; these pieces become the final state hadrons. In the
singlet cluster model, one imagines that each gluon splits into a
quark-antiquark pair and one then finds low mass color-singlet
clusters composed of the resulting partons.  These clusters decay into
hadrons.  In each picture, there are adjustable parameters tuned to
help fit the hadronization to experiment.  In each picture, the
hadronization is based on an approximation to the color flow in the
perturbative process.  Presumably this attention to the color flow
accounts for some of the success of these schemes in fitting the data.

{\bf Parton showers}. An essential feature of the Monte Carlos is their
simulation of the decay of a far off-shell parton into two less far
off-shell partons, and the subsequent decay of these partons into
further partons, producing a parton shower.  This simulation uses a
small angle approximation to the tree level QCD graphs.  The
implementation of this showering for incoming partons by means of
so-called ``backward evolution'' is new since 1983, as is the
approximate accounting for quantum interference embodied in ``angular
ordering.''

The best modern Monte Carlo event generators contain a great deal of
information about QCD and soft multiparticle physics. They have become
so good that there is a danger that experimentalists may mistake them
for the Standard Model, forgetting that the programs actually contain
an essentially tree level approximation to the Standard Model, mixed
up with a of hadronization model that is tuned to experiment.

The Monte Carlos have become essential tools for a number of
purposes.  First, one can use a Monte Carlo to estimate Standard Model
backgrounds for searches for new physics, being careful to keep track
of where the approximations built into the program are good and where
they are not. This is especially important in planning for future
experiments. Second, one can use the programs to estimate the effects
of detector resolution and efficiencies. There are inherent
difficulties with this, since the Monte Carlo simulations are only
approximate, but the results should be reliable as long as the
correction applied to the raw data based on these simulations is
small. Third, Monte Carlo programs can be used to estimate
hadronization corrections to be applied to perturbative theory.
One simply compares results with ``hadronization'' turned on and off.
Evidently, this is a rather crude method, but in cases where
hadronization effects are too large to ignore it is better than
nothing. Finally, the Monte Carlos provide a way for experimentalists
to explore the relation of their data to the underlying physical
processes. Analytical calculations may provide a more accurate tool
for special purposes, but such calculations cannot compete with the
scope of the Monte Carlo method, which is able to simulate many
different processes at once and to simulate whole events instead of
just selected features of events.

What of the future?  In my estimate, the power of the Monte Carlo
method is potentially very great.  One can imagine making the whole
thing work at the N-loop level instead of the 0-loop (tree) level,
putting more angular and spin information into the simulations,
developing better descriptions of the hadronization, and so forth.
Thus in 2003, the path theorists take between the fundamental
Lagrangian and experimental predictions may always be paved with the
blocks of a Monte Carlo program.

I have spoken of the use of Monte Carlo event generators to describe
multiparticle dynamics in the case where there is a ``hard
interaction.''  But there has been some success and there is great
potential for understanding soft hadronic physics also.  The potential
is to model a very complicated process with hundreds of effective
degrees of freedom using a simple underlying dynamics. One must be
aware, of course, that the Monte Carlo method is adapted to the
description of classical stochastic processes, whereas Nature uses
quantum probabilities rather than classical probabilities.  Thus the
method may be applied if there is a choice of variables for which a
semiclassical description works.  These variables may be the
coordinates or the momenta of quarks and gluons; the coordinates or
the momenta of pions and rhos; the configurations of color flux tubes,
or other possibilities. In this view, the art of modelling hinges on
the choice of variables.

\vglue 0.6cm
{\bigrm\noindent 7. Parton distribution functions}
\vglue 0.4cm

Knowledge of parton distributions is essential for making predictions
in processes with one or two hadrons in the initial state.  One needs
this knowledge both for making predictions based on the Standard Model
and for predictions based on extensions to the Standard Model such as
supersymmetry.  Thus, for instance, in 1983 one needed to know the
distribution of quarks in the proton in order to predict the cross
sections for $W$ and $Z$ boson production.

Today, one determines the distribution functions for all the partons
at once in a global fit of data from many processes, as described by
J.~Morfin at this conference \cite{morfin}. In 1983, the science of
making such global fits was just beginning.  One had the fits of Buras
and Gaemers \cite{bg}.  In 1983 the first of the Duke-Owens fits was
produced \cite{do}, along with the fit of Eichten, Hinchliffe, Lane and
Quigg \cite{ehlq}.  In 1993, the available fits are much more accurate.
First, the data are better.  Second,the theory is better.
Next-to-leading order calculations are used for the theoretical cross
sections that are compared to the data used in the fit;
next-to-leading order parton evolution is used.  Currently there are
two families of parton distribution sets that are regularly updated by
their authors to take into account data as it becomes available.
These are the sets by Martin, Roberts, and Stirling \cite{mrs} and by
the CTEQ Collaboration \cite{cteq}.  We will see an example in the
following section of how better parton distributions have helped
improve predictions.

\vglue 0.6cm
{\bigrm\noindent 8. Jet production in hadron collisions}
\vglue 0.4cm

In 1982, the UA2 group demonstrated the existence of clearly
visible jets in proton-antiproton collisions at the CERN collider.
Jets had been seen and studied in electron-positron collisions, but it
had been uncertain whether they would be a useful tool for examining
quark and gluon interactions in the more complicated environment of
hadron collisions at the CERN collider energy. The following quote
from J.~Hansen, representing UA2 at the 1983 Multiparticle
Conference summarizes their observation \cite{UA2jets}. ``The cross
section for jets with a given transverse energy is much larger at the
$p\bar p$ collider than at the ISR.  At large energies jets are
clearly visible, there is no need for an elaborate algorithm to
separate jets from the background.  This is clearly seen in fig.\ 3a,
which is a LEGO-plot of the energy distribution in the $\phi,\theta$
plane for the event with the highest transverse momentum.'' Figure 3a
from that talk, not reproduced here, was a plot in the now-familiar
Lego format showing two distinct jets of about 80 GeV transverse
energy each.

\begin{figure}[htb]
\vspace{1in}
{\bf This figure was pasted in with glue, in the style of the
Dark Ages. It is not available in postscript. Sorry.\hfil}
\vspace{1in}
\hfill\parbox{6in}{
\caption{Inclusive jet production cross section presented by
the UA2 Collaboration at the 1983 Multiparticle Conference
\protect\cite{UA2jets}. The error bars include the statistical and the
$E_T$ dependent systematic errors.  One should add a 40\% $E_T$
independent systematic error.  The two curves A and B represent the QCD
prediction.}}\hfill
\label{figure1}
\end{figure}

At the 1993 Multiparticle conference both the UA1 and the UA2
groups gave talks on jet production.  I consider here the
simplest of their results, the one jet inclusive cross section
$d\sigma/d E_T d \eta$ to make a jet with transverse energy $E_T$ and
rapidity $\eta$, at $\eta = 0$. I reproduce in Fig.~1 the results
presented by UA2. The two curves A and B represent the QCD
prediction.  As you can see, the theoretical prediction has
an uncertainty of about a factor 3 either way from the midpoint
between the two curves.  This uncertainty was reported to arise from
the uncertainty in the parton distributions used in the calculation.

\begin{figure}[htb]
\vspace{1in}
{\bf This figure was pasted in with glue, in the style of the
Dark Ages. It is not available in postscript. Sorry.\hfil}
\vspace{1in}
\hfill\parbox{6in}{\caption{
Inclusive jet production cross section presented by the
CDF Collaboration at the 1983 Multiparticle Conference
\protect\cite{CDFjets}. The error bars include the statistical
errors only. The curve represents the next-to-leading-order QCD
prediction \protect\cite{eks}.}}\hfill
\label{figure2}
\end{figure}

Where are we now?  I display in Fig.~\ref{figure2} the result of the
CDF group for $d\sigma/d E_T d \eta$ averaged over a range of $\eta$
near $\eta = 0$ \cite{CDFjets}.  The systematic errors, not shown in
the figure, have been substantially reduced, to something like 30\%.
At the same time, the range of $E_T$ covered has substantially
increased. The plot also shows the prediction of QCD \cite{eks}.  The
largest theoretical uncertainty (not displayed) arises from the
uncertainly in the parton distribution functions (together with
$\alpha_s$), which is estimated \cite{eks} to be about 20\%.  This is
a graphic illustration of the improvement in our knowledge of parton
distribution functions.  There is also a substantial uncertainty from
truncating the perturbative cross section at a finite order of
perturbation theory.  In this case, the leading order and the
next-to-leading order contributions are included, and the uncertainty
from omitting next-to-next-to-leading order terms may be estimated to
be about 15\% \cite{eks}.  In the case of the 1982 curves, which were
leading order, one might have estimated a 50\% uncertainty from not
including next-to-leading order. In summary, both experiment and
theory have improved, so that one can now compare the two with a
combined uncertainty of perhaps 50\% over an $E_T$ range over which
the cross section changes by nine orders of magnitude.   Thus there
has now been a good chance for Nature to prove QCD wrong.

\pagebreak[3]
\vglue 0.6cm
{\bigrm\noindent 9. Lattice QCD}
\vglue 0.4cm

At the 1983 Multiparticle conference,  H.~Quinn presented a talk
titled ``A word of caution'' warning against overconfidence in
perturbative QCD. During the discussion period after the talk, the
following exchange occurred.

\begin{quote}
J. Rushbrook (Cambridge): For a specific theory to be useful, it has
to be falsifiable, i.e., we have to be able to know when it fails.
Does QCD satisfy that criterion?

Quinn: Not yet.

Rushbrook:  Will it ever?  Will we know that it will?

Quinn: I don't think we will know it from jet physics.  I think if we
know it we will know it because we keep trying to do some hard
calculations in QCD and get to the point that perhaps eventually we
will have non-perturbative methods that will calculate the hadron
spectrum in QCD.  After all it's supposed to be the fundamental
theory of hadrons; it ought to calculate a few fundamental
parameters, like the mass of the proton over the mass of the rho.
Those are the things that are the real tests of QCD.

\end{quote}

Today lattice QCD had made great strides. Much of the development has
been in the direction of calculating hadronic matrix elements of weak
decay operators.  However, I would like to focus on an example along
the lines suggested by Quinn: the calculation of hadron masses in the
valence approximation reported this year by  Butler, Chen, Sexton,
Vaccarino, and Weingarten \cite{weingarten}.

\begin{table}[htb]
\begin{center}
\begin{tabular}{|l|c|c|}\hline
&Calculated&Measured\\\hline
$M(N)/M(\rho)$ \rule{0mm}{4mm}& 1.219$\pm$0.105 & 1.222\\\hline
$M(\Phi)/M(\rho)$ \rule{0mm}{4mm}& 1.333$\pm$0.032 &
1.327\\\hline
$M(\Delta)/M(\rho)$ \rule{0mm}{4mm}& 1.595$\pm$0.111 & 1.604\\\hline
$M(\Omega)/M(\rho)$ \rule{0mm}{4mm}& 2.298$\pm$0.098 &
2.177\\\hline
$[M(\Xi)+ M(\Sigma)- M(N)]/M(\rho)$ \rule{0mm}{4mm}&
1.930$\pm$0.073 & 2.047\\\hline
\end{tabular}
\parbox{6in}{
\caption{Calculated values of ratios of hadron masses to the mass of
the $\rho$ meson, from Weingarten {\it et al.}
\protect\cite{weingarten}.  There are also predictions for
$M(K^*)/M(\rho)$, $M(\Sigma^*)/M(\rho)$, and $M(\Xi^*)/M(\rho)$,
but these may be viewed as being involved in the fitting for the
quark masses rather than being independent predictions. }}
\end{center}
\end{table}

In this calculation, we lack numerical control on how good the valence
approximation (no quark loops) is.  However, the other approximations
are controlled. For example, the simulated hadrons are in a universe
of finite size not much bigger than a hadron size; the distance
between neighboring lattice points is not zero; chiral symmetry is
broken by the lattice approximation and must be restored.  The authors
attempt to correct for these approximations and estimate the error
involved in doing so. Their results, with errors, are shown in
Table~4.  One should note that if the results did not match
experiment, we could blame the valance approximation instead of the
fundamental theory.  Nevertheless, the results go a long way toward
answering Quinn's challenge.

\vglue 0.6cm
{\bigrm\noindent 10. Multiparticle correlations}
\vglue 0.4cm

At the 1983 Multiparticle Conference, there was considerable
discussion of the distribution of particles produced in soft hadron
collisions (without high transverse momentum jets)
\cite{soft,dewolf}.  The participants were interested in the average
number of particles per unit rapidity, $dN/dy$, the probability to
get a given total multiplicity, $P(N)$, correlations between the
numbers of particles with large positive rapidity and large negative
rapidity $y$, the distribution of electric charge as a function of
rapidity, and so forth.  Models such as the Dual Parton Model
\cite{dpm} were proposed to explain the data, while too-simple models
were ruled out. However, most models survived, as reflected in the
following quote from the introduction to the talk of E.\ A.\ De Wolf
\cite{dewolf}. ``Many models have been proposed and were found to be
overall quite successful.  Because of large differences in their
dynamical input at the `microscopic' level \dots it has come as a
surprise that essentially all models agree with most of the data
examined.''

In the years since 1983, the examination of multiparticle
distributions has become more exacting.  Thus at the 1993 conference
there was much emphasis on examining the correlations between
particles with small separations in $y$ and $\phi$.
One is finding that correlations exist on all scales of $\Delta y$,
$\Delta \phi$ down to quite small values. A favored method for looking
for these features of the data uses intermittency analysis \cite{bp},
but other measures were discussed as well.

The remarkable fact, which serves as an indication of progress, is
that this increasingly sophisticated experimentation and data analysis
is proving capable of challenging the models.  Thus, for instance,
the Dual Parton Model, although it gets the gross features of
multiparticle production right, does not correctly reproduce the
intermittency behavior displayed by the data \cite{dpmwrong}.

\vglue 0.6cm
{\bigrm\noindent 11. Rapidity gap physics}
\vglue 0.4cm

I now turn to a subject that is both old and new, rapidity gaps in
scattering events.  This subject can be introduced by considering
elastic scattering, a venerable topic that was discussed at the 1983
multiparticle conference.  We consider elastic scattering of two
hadrons at large $s$, small $t$, as depicted in Fig.~\ref{figone}.
(Diffractive dissociation events are similar, but I do not consider
them here.) A plot of transverse energy deposition in the calorimeter
versus azimuthal angle $\phi$ and (pseudo-)rapidity $\eta$ for elastic
scattering is very simple, as shown in Fig.~\ref{figtwo}.  Hadron A
appears in the final state at large positive rapidity $\eta \approx
{1\over 2}\ln(s/|t|)$, while hadron B appears at large negative
rapidity $\eta \approx -{1\over 2}\ln(s/|t|)$. In a typical hadron
scattering event, the calorimeter is full of particles covering the
whole range of pseudorapidities.  In an elastic scattering event,
there is a large gap in the rapidity space in which there are no
particles.

\begin{figure}[htb]
\centerline{\DESepsf(aspen1.epsf width 2.0 in)}
\caption{Elastic scattering.}
\label{figone}
\end{figure}

\begin{figure}[htb]
\centerline{\DESepsf(aspen2.epsf width 3.2 in)}
\caption{Elastic scattering transverse energy deposition
versus $\eta$ and $\phi$.}
\label{figtwo}
\end{figure}

The quantum system exchanged between the hadrons in elastic scattering
at large $s$ and small $t$, represented by the jagged line in
Fig.~\ref{figone}, is called the pomeron. There was considerable
discussion of the nature of the pomeron at this conference
\cite{pomeron93}. The best current analyses involves sums of diagrams
like that shown in Fig.~\ref{figthree}, together with more complicated
diagrams. In this picture, I have endeavored to depict the time
evolution, with gluons emitted from the hadrons long before the
scattering and being absorbed into the hadron state long afterwards.
The theoretical questions being addressed in current studies of the
pomeron are difficult, and it is not easy to follow the details, but
the physical picture of interacting gluon clouds, as illustrated in
Fig.~\ref{figthree}, may be helpful as a guide to understanding.

\begin{figure}[htb]
\centerline{\DESepsf(aspen3.epsf width 4.4 in)}
\caption{Typical graph contributing to the pomeron.}
\label{figthree}
\end{figure}

The simplest perturbative version of pomeron exchange is the
scattering of two partons by exchange of two hard gluons, as depicted
in Fig.~\ref{figfour}.  Here we ask that $\hat s \gg |\hat t|$ and
that the exchanged gluons form a color singlet.  We also ask that
$\hat t$ be large compared to $1 {\rm \ GeV}^2$, and indeed that {\em
both} gluons carry large transverse momentum, so that we are
describing a short distance process. If, in a hadron-hadron
scattering event, two partons scatter as in Fig.~\ref{figfour}, the
outgoing partons will show up as high $p_T$ jets, separated in
rapidity by   $\Delta\eta \approx \ln(\hat s/|\hat t|)$. If nothing
else happened in the event, there would be no particles produced in
the gap between the two jets.  Bjorken has argued \cite{bjgap} that
even allowing for the other things that can happen, there is still a
not-tiny probability for this gap to survive. (If this is so, then
there will be applications for other hard scattering processes with
color singlet exchange, such as Higgs boson production at the SSC.)
One process that is sure to happen is gluon bremsstrahlung from the
scattered partons.  One expects collinear radiation along the beam
directions and along the jet directions.  One also expects soft
radiation into the regions between the jets and the beams.  That is,
if the forward parton is scattered to angles $(\eta_1,\phi_1)$, then
one expects radiation near $\eta = \eta_1$, $\phi = \phi_1$, near
$\eta = \infty$, and in the range $\eta_1 < \eta < \infty$.
Similarly, there will be radiation near $\eta = \eta_2$, $\phi =
\phi_2$, near $\eta = -\infty$, and in the range $ - \infty< \eta <
\eta_1$.  However as long as the scattering was accomplished by a
hard color singlet exchange, there is no bremsstrahlung in $ \eta_2
\ll \eta \ll \eta_1$. The gap remains. One must now ask if soft
collisions between the remaining quarks from the two hadrons will
produce particles in the whole calorimeter, thus spoiling the gap.
The probability that spectator collisions do not fill the gap is
tentatively estimated as a few percent \cite{bjgap}. The event
structure for events with surviving gaps is illustrated in
Fig.~\ref{figfive}.

\begin{figure}[htb]
\centerline{\DESepsf(aspen4.epsf width 2.5 in)}
\caption{Elastic scattering of two partons by two gluon exchange.}
\label{figfour}
\end{figure}

\begin{figure}[htb]
\centerline{\DESepsf(aspen5.epsf width 3.2 in)}
\caption{Transverse energy deposition for partonic elastic
scattering with a gap.}
\label{figfive}
\end{figure}

At this meeting, the D0 group at the Fermilab Collider reported
seeing events with the signature indicated in Fig.~\ref{figfive}
\cite{D0gap}.

\begin{figure}[htb]
\centerline{\DESepsf(aspen6.epsf width 4.4 in)}
\caption{Normal deeply inelastic scattering.}
\label{figsix}
\end{figure}

There is another kind of gap event that was anticipated for deeply
inelastic scattering events at HERA. However, here the underlying
physics is quite different.  Consider a deeply inelastic electron
scattering event in which $x_{bj} \ll 1$.  A typical Feynman diagram
representing such an event is shown in Fig.~\ref{figsix}. The struck
quark is scattered through a large angle and emerges as the ``
current jet'' with rapidity $\eta_J$. The proton remnants have large
positive rapidity and produce particles in the calorimeter near $\eta
= \infty$. This struck quark, which carries a small fraction $x_{bj}$
of the proton's momentum, is produced by the successive splitting of
partons carrying larger momentum fractions. In the diagram, the
partons enter the final state. They fill the calorimeter at
rapidities between $\eta_J$ and $\infty$.  However, it is possible
for most of the gluons in the diagram to recombine with the valence
quarks of the proton and thus reconstitute the proton, as indicated
in Fig.~\ref{figseven}. One might call this recombinant
bremsstrahlung. The proton enters the final state having lost a small
fraction ($> x_{bj}$)  of its longitudinal momentum and having a
small amount of transverse momentum transferred to it. This
phenomena, called diffractive hard scattering, or diffractive deeply
inelastic scattering in this case, was anticipated by Ingelman and
Schlein \cite{is}.  As is suggested by the diagrams, the phenomenon
is similar to elastic scattering, but with the pomeron probed by the
hard virtual photon.

\begin{figure}[htb]
\centerline{\DESepsf(aspen7.epsf width 4.4 in)}
\caption{Diffractive deeply inelastic scattering.}
\label{figseven}
\end{figure}

In diffractive deeply inelastic scattering, one should see the
elastically scattered proton at large rapidity but with a large gap
around $\eta = \infty$, as shown in Fig.~\ref{figeight}.

\begin{figure}[htb]
\centerline{\DESepsf(aspen8.epsf width 3.2 in)}
\caption{Transverse energy deposition for diffractive deeply
inelastic scattering.}
\label{figeight}
\end{figure}

At this meeting, the Zeus and H1 groups at HERA reported
seeing events with gaps as indicated in Fig.~\ref{figeight}
\cite{HERAgap}.

\begin{figure}[htb]
\centerline{\DESepsf(aspen9.epsf width 4.4 in)}
\caption{Diffractive jet production.}
\label{fignine}
\end{figure}

Diffractive hard scattering can also been seen in hadron-hadron
collisions. Here the hard scattering is parton-parton scattering to
produce jets.  If the jets are produced by collision of a small $x$
parton from one of the hadrons, then the partons that carry most of
that hadron's momentum can recombine so that the hadron appears again
in the final state, having lost a small fraction $z$ of its
longitudinal momentum. This is depicted in Fig.~\ref{fignine}.  Again,
there is a gap, as sketched in Fig.~\ref{figten}.  This type of event
was predicted in \cite{is} and seen by the UA8 experiment at
the CERN collider \cite{UA8}.

\begin{figure}[htb]
\centerline{\DESepsf(aspen10.epsf width 3.2 in)}
\caption{Transverse energy deposition for diffractive jet
production.}
\label{figten}
\end{figure}

I would like to point out a feature of hard diffractive scattering
that I am currently studying with A. Berera \cite{bs}.  In
Fig.~\ref{fignine}, one gluon from the pomeron participates in the
hard scattering, while one enters the final state.  Of course, more
than one gluon could have been emitted.  There must be at least one
``extra'' gluon because the pomeron carries zero color charge, while
the gluon that participates in the hard scattering is a color octet.
The extra gluon in Fig.~\ref{fignine} carries away the extra color
charge, and it also carries away some longitudinal momentum.  Thus the
momentum fraction $x$ delivered to the hard interaction is less than
the momentum fraction $z$ carried by the pomeron.  However, there is
another possibility.  The extra color can be transferred to the
spectator partons of the oppositely moving hadron as shown in
Fig.~\ref{figeleven}.  The gluon exchanged in this process carries
negligible longitudinal momentum.   Such graphs exist in normal hard
processes such as jet production, but their effect cancels because of
unitarity.  Here the unitarity argument does not apply because one is
not summing over all final states, but rather is demanding that the
initial hadron emerges again, slightly scattered, in the final state.
This sort of violation of the normal hard scattering factorization
has also been studied by Collins, Frankfurt, and Strikman
\cite{cfs}.  There is some evidence in the experimental results of
UA8 \cite{UA8} that for many of the events $1 - x/z$ is small,
indicating either that the ``extra'' gluon(s) in Fig.~\ref{fignine}
are often quite soft or that the graphs like those in
Fig.~\ref{figeleven} are important.  In either case, it would appear
that the same effect should {\em not} be seen in diffractive deeply
inelastic scattering: in deeply inelastic scattering the ``extra''
parton is a quark, which does not like to be soft.

\begin{figure}[htb]
\centerline{\DESepsf(aspen11.epsf width 4.4 in)}
\caption{Diffractive jet production with color transferred to
spectator parton.}
\label{figeleven}
\end{figure}

\vglue 0.6cm
{\bigrm\noindent 12. Conclusion}
\vglue 0.4cm

I have tried to compare where we are now in the study of multiparticle
dynamics compared to where we were at the time of the 1983
conference.  Naturally, I  have only been able to touch on some
selected topics for this comparison.  Nevertheless, I have personally
found it to be an instructive exercise.  I have found the evolution
of the field to be a little like a hike in the mountains
surrounding Aspen.  It often seems as though you are moving
slowly, considering how far there is to go, but if you stop to look
back you can see that you have covered more ground than you thought.

\vglue 0.6cm
{\bigrm\noindent Acknowledgements}
\vglue 0.4cm

I would like to thank  R.\ Hwa, J.\ Morfin, and T.\ Sjostrand for
providing information that was useful in the preparation of this talk.
This work was supported in part by U.S. Department of Energy grant
DE-FG06-85ER-40224.
\pagebreak[3]

\vglue 0.6cm
{\bigrm\noindent  References}
\vglue 0.4cm

\nopagebreak

\end{document}